\documentclass[twocolumn]{aastex631}
\usepackage{CJK}
\usepackage{natbib}
\usepackage{graphicx}
\usepackage {amsmath,amssymb,eqnarray}
\usepackage{enumerate}
\usepackage{enumitem}
\usepackage{mathtools}

\shorttitle{Measurement of splashback radius}
\shortauthors{Xu et al.}

\begin{document}
\title{The measurement of the splashback radius of dark matter halo}

\author[0000-0002-9587-6683]{Weiwei Xu}
\thanks{wwxu@pku.edu.cn}
\affiliation{National Astronomical Observatories (NAOC), Chinese Academy of Sciences, Beijing 100101, China}
\affiliation{Institute for Frontiers in Astronomy and Astrophysics, Beijing Normal University, Beijing 102206, China}
\affiliation{The Kavli Institute for Astronomy and Astrophysics, Peking University (KIAA-PKU), Beijing 100871, China}
\author[0000-0001-8534-837X]{Huanyuan Shan}
\thanks{hyshan@shao.ac.cn}
\affiliation{Shanghai Astronomical Observatory (SHAO), Nandan Road 80, Shanghai 200030, China}
\affiliation{Key Laboratory of Radio Astronomy and Technology, Chinese Academy of Sciences, A20 Datun Road, Chaoyang District, Beijing, 100101, P. R. China}
\affiliation{University of Chinese Academy of Sciences, Beijing 100049, China}
\author[0000-0003-3899-0612]{Ran Li}
\affiliation{National Astronomical Observatories (NAOC), Chinese Academy of Sciences, Beijing 100101, China}
\affiliation{Institute for Frontiers in Astronomy and Astrophysics, Beijing Normal University, Beijing 102206, China}
\affiliation{School of Astronomy and Space Science, University of Chinese Academy of Science, Beijing 100049, China}
\author[0000-0002-7336-2796]{Ji Yao}
\affiliation{Shanghai Astronomical Observatory (SHAO), Nandan Road 80, Shanghai 200030, China}
\author{Chunxiang Wang}
\affiliation{National Astronomical Observatories (NAOC), Chinese Academy of Sciences, Beijing 100101, China}
\affiliation{Institute for Frontiers in Astronomy and Astrophysics, Beijing Normal University, Beijing 102206, China}
\affiliation{School of Astronomy and Space Science, University of Chinese Academy of Science, Beijing 100049, China}
\author[0000-0001-6800-7389]{Nan Li}
\affiliation{National Astronomical Observatories (NAOC), Chinese Academy of Sciences, Beijing 100101, China}
\author[0000-0002-3286-9099]{Chaoli Zhang}
\affiliation{College of Computer Science and Artificial Intelligence, Wenzhou University, 325035 Wenzhou, China}

\begin{abstract}

In the hierarchical evolution framework of cosmology, larger halos grow through matter accretion and halo mergers. To clarify the halo evolution, we need to define the halo mass and radius physically. However, the pseudo-evolution problem makes the process difficult.
Thus, we aim to measure the splashback radius, a physically defined halo radius for a large number of halos with various mass and redshift,
and to determine the most important parameters to affect it. 
We use the typical definition of splashback radius as the radius with the steepest radial density profile.
In this work, we measure the splashback radius ($R_{\rm sp}$) of dark matter halos within the mass of $10^{13}{\rm M}_{\odot}$ to $3\times10^{15}{\rm M}_{\odot}$ and redshifts spanning $0.08$ to $0.65$. This is the measurement of the $R_{\rm sp}$ in the largest range of halo mass and redshift. Using the shear catalog of the DECaLS DR8,
we investigate the splashback radius of halos associated with galaxies and galaxy clusters identified in the various catalogs.
Our finding reveals a trend wherein massive halos demonstrate a larger splashback radius, and the normalized splashback radius ($R_{\rm sp}$/$R_{\rm 200m}$) shows a U-shaped mass evolution. 
The upturn in these relations mainly comes from the contribution of massive halos with low redshifts. We further find the splashback radius increases with the peak height, while the normalized splashback radius has a negative relation with the peak height. We also find the $R_{\rm sp} \gtrsim R_{\rm 200m}$ for most halos, indicating their low accretion rates.
Our result is consistent with previous literature across a wide range of mass, redshift, and peak height, as well as the simulation work from \cite{More2015}.
\end{abstract}

\section{Introduction}

In the standard $\Lambda$CDM cosmological framework, cosmic structures grow through a hierarchical process \citep{Gunn1972}. The formation of substantial dark matter halos originates from the initial high-density peak in the early universe \citep{Fillmore1984}. As matter accumulates in the region, a deep gravitational well forms, performing a large attraction to nearby materials. Finally, a large dark matter halo forms. Once the outer boundary of the dark matter halo is well-defined, the halo mass is constrained, which is the most important parameter of the dark matter halo. 

A natural question arises regarding how to define the outer boundary of the dark matter halo. Commonly employed radii include the virial radius and the radii defined with the overdensity, such as $R_{500}$ and $R_{200}$. Nonetheless, these radii likely suffer from the pseudo-evolution of the halo mass, as demonstrated in \cite{Diemer2013}. This phenomenon results in two identical halos located at different redshifts appearing to possess varying masses, due to the redshift evolution of the cosmological critical density. Consequently, it is vital to establish a physically-motivated boundary of the halo to accurately characterize the evolution of the dark matter halo \citep{Diemer2014, Adhikari2014, More2015}.
Furthermore, the dependence of assembly bias on halo mass definitions is closely linked to the radius definition, due to the dependence of halo clustering on halo property other than mass \citep{More2016, Villarreal2017, Chue2018, Mansfield2020}. 

The splashback radius ($R_{\rm sp}$) is one of the physically motivated radii of dark matter halo.
Within the framework of the self-similar spherical collapse model, the dark matter halos can be characterized by a set of infinitesimally thin mass shells. These shells are initially coupled with the Hubble flow, undergoing accelerated expansion before decelerating, turning around, and collapsing into the halo, and eventually virializing under the influence of gravitational forces \citep{Fillmore1984, Bertschinger1985}. The splashback radius is thus defined as the boundary separating virialized and infalling shells. 
Particle apocenter scattering \citep{Adhikari2014, Mansfield2017} implies that halos and the splashback radii may not be perfectly spherical, and it is also influenced by the energy and momentum of particles as they begin to infall \citep{Diemer2017}. In the simulation, the splashback radius is estimated with the apocenter of the accreted dark matter particles in their first orbit motion \citep{Diemer2017}. The work of \cite{Pizzardo2023} also confirms with IllustrisTNG simulation that the splashback radius effectively labels the inner boundary of the region where infalling matter dominates.

For simplicity, the location with the steepest slope of 3D density profile $\rho(r)$ is usually taken as the splashback radius, especially in observations (e.g., \citealt{Diemand2008, Cuesta2008, More2015}). In \cite{More2015}, the small radial velocity near the apocenter results in the accreted matter piling up in the nearby region, and further the density enhancement caused a \rm{caustic} in the splash-back area. Thus, the splashback radius, defined as the orbital apocenter of the accreted matter, corresponds to the \rm{outermost} \rm{caustic}. In the spherical collapse model, the halo within the splashback radius includes approximately all material ever accreted, thus the change of $M_{\rm sp}$ comes from the new accreted matter, which is barely affected by the pseudo-evolution.

Using the splashback radius as a proxy for the halo boundary, it is possible to distinguish discrepancies in the assembly bias measurements \citep{Chue2018}. Additional researches \citep{Adhikari2016, Adhikari2018, Banerjee2020} indicate that the splashback radius may serve as a tool for assessing dynamical friction and constraining alternative theories for gravity and self-interacting dark matter. 
Apart from the splashback radius, other radii have been proposed for various purposes.
For instance, Fong et al. \citep{Fong2021, Fong2022} have employed the depletion radius relating to the halo bias and infalling velocity to describe the depleted area of growing halos, which is approximately $2.5$ times of the virial radius and $1.7–3$ times of the splashback radius.

Many works have been dedicated to measuring the splashback radius. 
The first detection of the splashback radius was made by \cite{More2016}. They measure the surface density profile around redMaPPer \citep{Rykoff2014} clusters from the Sloan Digital Sky Survey (SDSS) Data Release 8 (DR8), and find the model with splashback radius \citep{Diemer2014} fits the profile better than the model without this feature.
Then, in the work of \cite{Baxter2017}, redMaPPer clusters display an abrupt decrease of the fraction of red galaxies around the location of the splashback radius. 
In \cite{Chang2018}, they measure the splashback radius with the galaxy number density and weak lensing mass profiles of the redMaPPer clusters from first-year Dark Energy Survey (DES) data. They use the same method with the work of \cite{More2016} and \cite{Baxter2017}, and obtain consistent results with them. Their measurements are also consistent with $\Lambda$CDM simulations. In addition, they conclude the $r_{\rm sp}$ scales with $R_{\rm 200m}$ and it does not evolve with redshift within the redshift of $0.3–0.6$. Then, the splashback radius is detected for more optically selected clusters \citep{Murata2020, Bianconi2021, Contigiani2023}, Sunyaev-Zel'dovich effect(SZ)-selected clusters \citep{Shin2019, Shin2021, Zurcher2019}, intra-cluster light of clusters \citep{Deason2021}.
In addition, multiple works investigate splashback radius in simulations (such as \citealt{Xhakaj2020}).

In multiple works (e.g., \citealt{Diemer2014, More2015, Chang2018}), the splashback radius is typically defined or measured as the radius with the steepest density slope (called $R_{\rm steep}$ in this paragraph for convenience). However, some recent works (e.g., \citealt{Aung2021, Diemer2022, Garcia2023}) find the splashback radius is not necessarily the same with $R_{\rm steep}$, especially for less massive halos. In less massive halos, splashback features are weaker because of their lower mass accretion rates on average. Besides, their splashback features occur at a larger radius, where the infalling term becomes important or even dominates. In this situation, $R_{\rm steep}$ is the radius where the steep inner profile and shallow outer profile trade-off. However, to keep consistent with previous relevant works, we keep the typical way to define and characterize the splashback radius, as $R_{\rm sp}=R_{\rm steep}$, in this work.

In this paper, we measure the halo splashback radius with the galaxy-galaxy lensing method, encompassing halo mass between $10^{13}$ and $3\times10^{15}~{\rm M}_{\odot}$ and a redshift range of $0.08-0.65$. Throughout the paper, we adopt M$_{200\rm m}$ and $c_{200\rm m}$ as the mass and concentration of the halo and use the comoving distance.

The cosmological parameters used in this paper are obtained from \citet{Planck2018}, 
consisting of the Hubble constant $H_0=67.4$~km~s$^{-1}$~Mpc$^{-1}$, 
baryon density parameter $\Omega_{\rm b}h^2=0.0224$, 
cold dark matter density parameter $\Omega_{\rm cdm}h^2=0.120$, 
matter fluctuation amplitude $\sigma_8=0.811$, 
power index of primordial power spectrum n$_{\rm s}=0.965$,
and matter density parameter $\Omega_{\rm m} = 0.315$.
This paper is organized as follows: 
Sec.~\ref{sec:data} introduces the source and lens catalogs;
Sec.~\ref{sec:method} presents the lensing signal, lensing model, and associated systematics; 
Sec.~\ref{sec:result} shows and discusses the results. 
We describe the conclusion in Sec.~\ref{sec:conclusion}.

\section{Data}
\label{sec:data}
\subsection{Source catalog}
\label{subsec:source_catalog}

In our measurement of the lensing signal, we use the shear catalog from the Dark Energy Camera Legacy Survey Data Release 8 catalog (DECaLS DR8) covering a sky area of 9,500 deg$^2$, as a part of the Dark Energy Spectroscopic Instrument Legacy Imaging Survey \citep{Blum2016, Dey2019}. The shear catalog is obtained mainly with the following steps.
\begin{itemize}
  \item In the shear measurement, five morphologies are used to describe and divide the {\it Tractor} source \citep{Lang2014}. The models include point sources, round exponential galaxies with a variable radius, DeVaucouleurs, exponential, and the composite model. The DECaLS DR8 sources below the $6\sigma$ detection limit in all stacks are removed. 
  \item A joint fitting on the optical images (in $g$, $r$, and $z$ bands) is used to estimate the ellipticity of a galaxy. 
  \item We estimate the photometric redshift of galaxy with the $k$-nearest-neighbour algorithm (kNN, \citealt{Zou2019}). For reliability, we use not only the three optical bands ($g$, $r$, and $z$), but also two infrared bands ($W1$ and $W2$) from the Wide-Field Infrared Survey Explorer, and only reserve the galaxies with $r<23$~mag. This way, we obtain the catalog of about $2.2$ million galaxies with photometric redshifts. 
\end{itemize}

The performance of the final catalog \citep{Zou2019} includes the redshift bias of $\Delta z_{\rm norm}=2.4\times10^{-4}$, accuracy of $\sigma_{\Delta z_{\rm norm}}=0.017$, and outlier rate of $\sim5.1\%$. 

\subsection{Lens catalogs}

\begin{figure*}[t]
    \centering
    \includegraphics[width=0.3\textwidth]{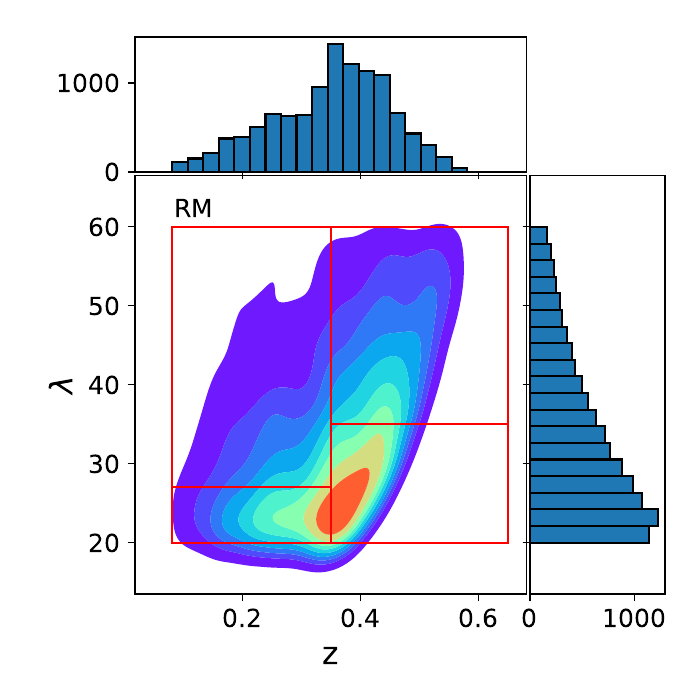}
    \includegraphics[width=0.3\textwidth]{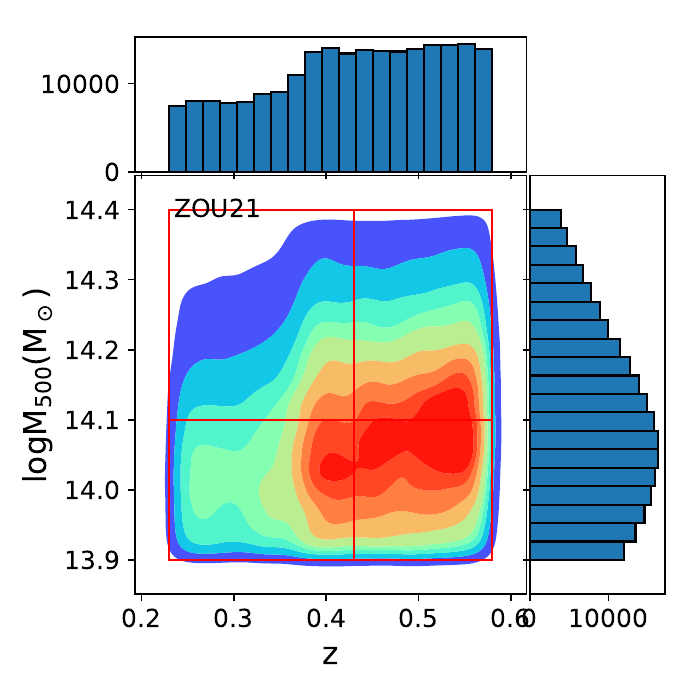}
    \includegraphics[width=0.3\textwidth]{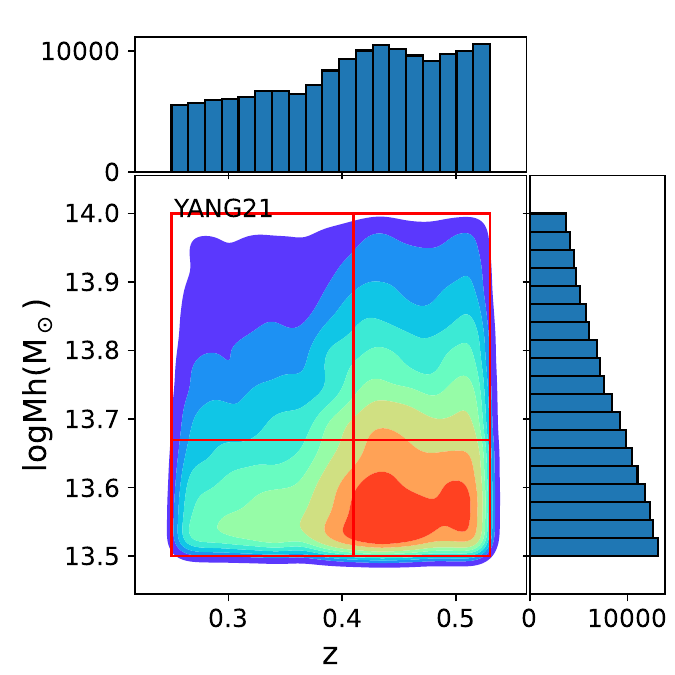} 
    \includegraphics[width=0.3\textwidth]{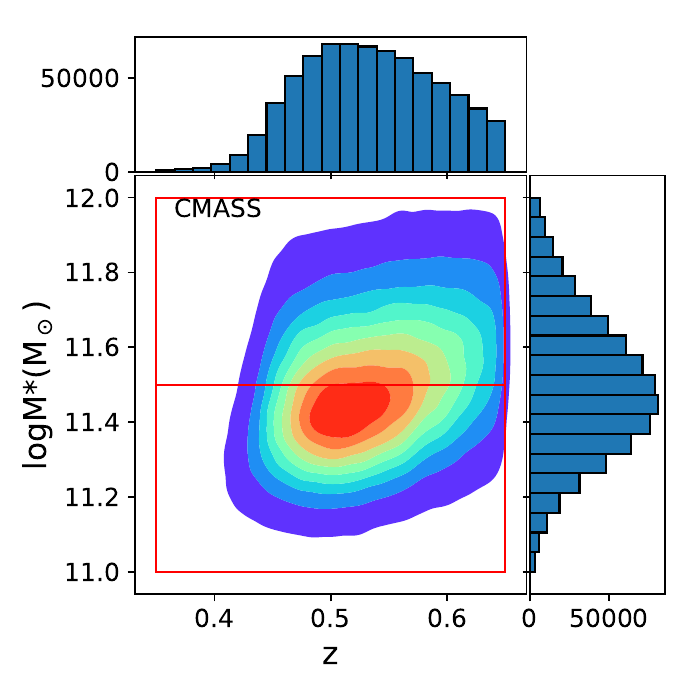}
    \includegraphics[width=0.3\textwidth]{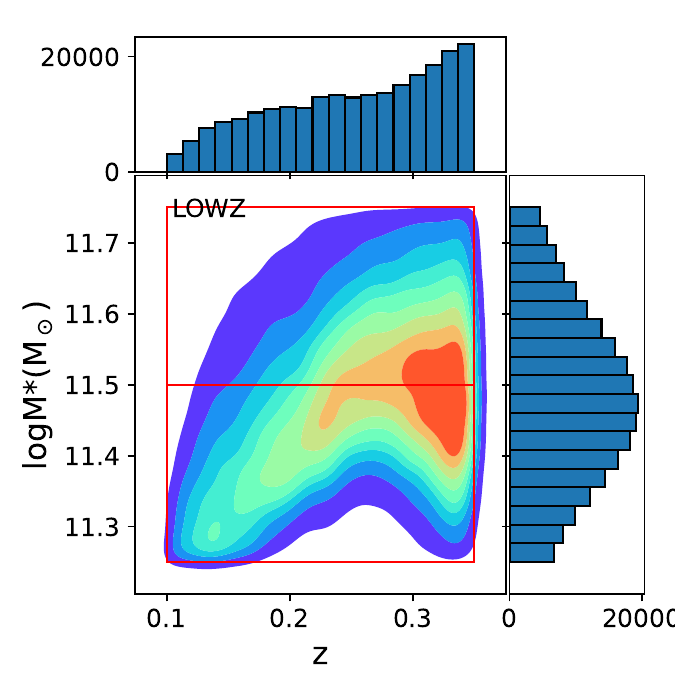}
    \caption{The bin criteria of lens catalogs in the redshift and mass proxy parameter space. From top to bottom, from left to right, the panels correspond to redMaPPer, ZOU21, YANG21, CMASS, and LOWZ catalog, in sequence. In each panel, the main plot illustrates the halo distribution in the redshift-mass space, with color representing halo number density. Redshift and mass proxy histograms are displayed in small plots above and to the right. Red horizontal and vertical lines indicate mass proxy and redshift thresholds. Galaxy clusters and groups beyond the labeled ranges are taken as outliers, and excluded from plots.}
    \label{fig:bins}
\end{figure*}

Halos from five catalogs are used as lenses, including the redMaPPer ('RM' for short, \citealt{Rykoff2014}), ZOU21 \citep{Zou2021}, YANG21 \citep{Yang2021}, CMASS \citep{Ahn2014}, and LOWZ \citep{Ahn2014} catalogs.
To constrain the effects of redshift and halo mass,
we divide each halo catalog into multiple 
bins, as shown in Fig.~\ref{fig:bins} and Tab.~\ref{tab:bins}. To accurately characterize the entire sample and avoid bias from outliers in the redshift and mass parameter space, we discard halos outside of listed parameter ranges. The mass proxy used is the richness for redMaPPer catalog, the logarithm of M$_{500}$ for ZOU21 catalog, the logarithm of halo mass for YANG21 catalog, and the logarithm of stellar mass for CMASS and LOWZ catalogs.

\begin{table}[t]
    \centering
    \caption{The bin criteria of lens samples.}
    \footnotesize
    \begin{tabular}{c|c|c|c|c}
    \hline  
    \hline  
    Cat.   & bin     &$z$              & Mass proxy   & Num.$^*$      \\ 
    \hline 
        RM & z1m1 & $0.08 - 0.36$ & $20 - 34$  &   $2,484$ \\
           & z1m2 & $0.08 - 0.36$ & $34 - 60$  &   $2,474$ \\
           & z2m1 & $0.36 - 0.65$ & $20 - 39$  &   $3,130$ \\
           & z2m2 & $0.36 - 0.65$ & $39 - 60$  &   $3,031$ \\  
    \hline  
   ZOU21   & z1m1 & $0.23 - 0.43$ & $13.90 - 14.10$ &  $57,998$  \\
           & z1m2 & $0.23 - 0.43$ & $14.10 - 14.40$ &  $49,119$  \\
           & z2m1 & $0.43 - 0.58$ & $13.90 - 14.10$ &  $55,704$  \\
           & z2m2 & $0.43 - 0.58$ & $14.10 - 14.40$ &  $58,365$  \\
    \hline
   YANG21  & z1m1 & $0.25 - 0.41$ & $13.50 - 13.67$ & $36,270$  \\
           & z1m2 & $0.25 - 0.41$ & $13.67 - 14.00$ & $36,646$  \\
           & z2m1 & $0.41 - 0.53$ & $13.50 - 13.67$ & $39,502$  \\
           & z2m2 & $0.41 - 0.53$ & $13.67 - 14.00$ & $41,848$  \\
    \hline
    CMASS  & m1 & $0.35 - 0.65$ & $11.00 - 11.50$ & $378,884$  \\
           & m2 & $0.35 - 0.65$ & $11.50 - 12.00$ & $338,123$  \\
    \hline
    LOWZ   & m1 & $0.10 - 0.35$ & $11.25 - 11.50$ &  $133,934$  \\
           & m2 & $0.10 - 0.35$ & $11.50 - 11.75$ &  $103,829$  \\
    \hline  
    \hline  
    \end{tabular}\newline
    {\bf Note.} {The first two columns are the catalog name and the bin name. In the bin name, the `m1' and `m2' represent less and more massive bins, `z1' and `z2' for low and high redshift bins, respectively. The thresholds of the redshift and mass proxy are listed in the 3rd and 4th columns. The last column exhibits the total number of halos included. The last column lists the total halo number in the specific parameter range, with halos located outside of DECaLS DR8 coverage also included.}
    \label{tab:bins}
\end{table}

We use the $26,111$ redMaPPer clusters (catalog v6.3\footnote{http://risa.stanford.edu/redmapper/}) derived from the SDSS DR8 \citep{Rykoff2014} as the lens catalog. The richness of these clusters are $\lambda>19$. In each cluster, we take the galaxy with the highest center likelihood as the halo center. 
For an accurate center determination, we only use clusters exhibiting high centering probabilities (P$_{\rm cen}>0.95$). 
To understand the relation between the splashback radius and other parameters, such as redshift and mass, the redMaPPer cluster sample is divided into four bins, with high$/$low redshift and richness. 

We divide redMaPPer clusters into high redshift and low redshift subsamples. Then, each subsample is separated into the less massive and more massive cluster bins. The redMaPPer halo bins are shown in the upper-left panel of Fig.~\ref{fig:bins}. Using the galaxy-galaxy lensing measurements, we obtain the surface density profiles of clusters with different halo mass and redshift. With similar halo numbers in each bin, we control potential biases from the sample size in the signal measurement.

In addition, we use the halo catalogs of ZOU21 \citep{Zou2021} and YANG21 \citep{Yang2021}. They identified $540,432$ galaxy clusters and $6.4$ million galaxy groups from the DESI legacy imaging surveys (the Legacy Surveys) with a fast clustering algorithm and luminosity function, respectively. For the ZOU21 catalog, the density peak is taken as the halo center. For the YANG21 catalog, we use the group catalog obtained from the DESI Image Legacy Surveys DR9 data\footnote{https://gax.sjtu.edu.cn/data/DESI.html} and take the position of the most massive member galaxy as the halo center. For reliability, we only use YANG21 groups with richness $>5$. Both ZOU21 and YANG21 catalogs are divided into four bins, for less massive and more massive halos at low redshift and high redshift. The distributions of halos are shown in the upper-middle and upper-right panel of Fig.~\ref{fig:bins}.

Besides, we also utilize the CMASS and LOWZ catalogs from SDSS-III BOSS DR10 \citep{Ahn2014} to constrain the splashback radius at low mass regions. The redshifts of CMASS and LOWZ halos are constrained within $0.35-0.65$ and $0.10-0.35$, respectively. For these two catalogs, we take the stellar mass as the halo mass proxy and separate each catalog into two mass bins, with the distributions shown in the second row of Fig.~\ref{fig:bins}. 

\section{Method}
\label{sec:method}
\subsection{Measurement of galaxy-galaxy lensing signal}

\begin{figure*}[t]
    \centering
    \includegraphics[width=0.3\textwidth]{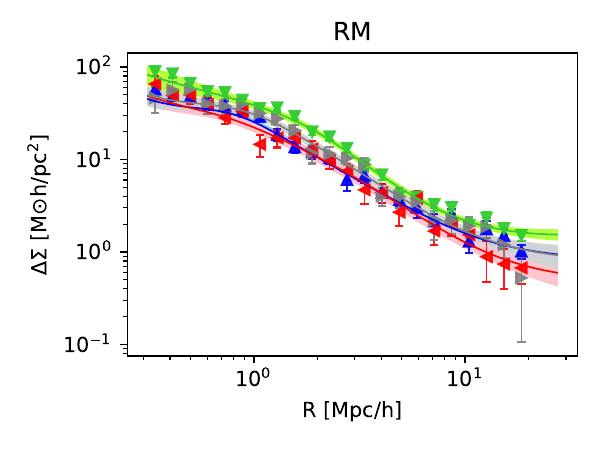}
    \includegraphics[width=0.3\textwidth]{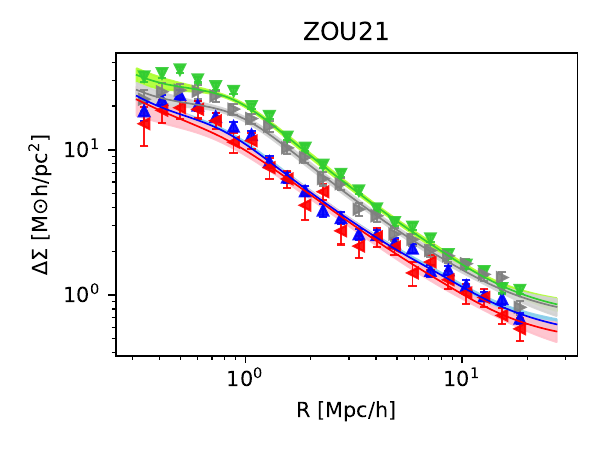}
    \includegraphics[width=0.3\textwidth]{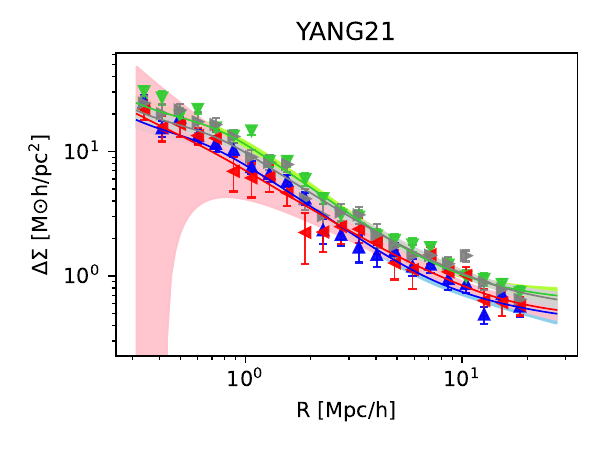}
    \includegraphics[width=0.3\textwidth]{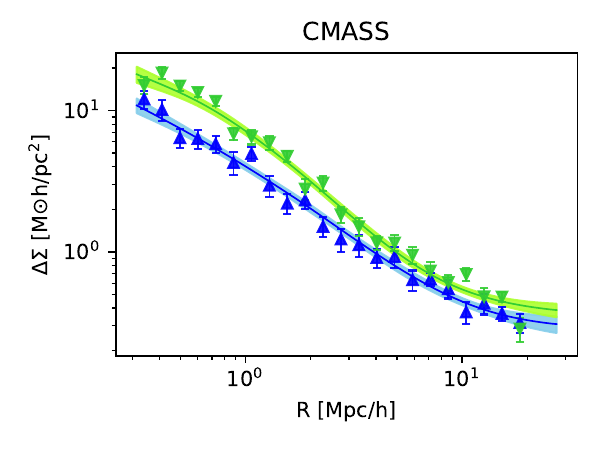}
    \includegraphics[width=0.3\textwidth]{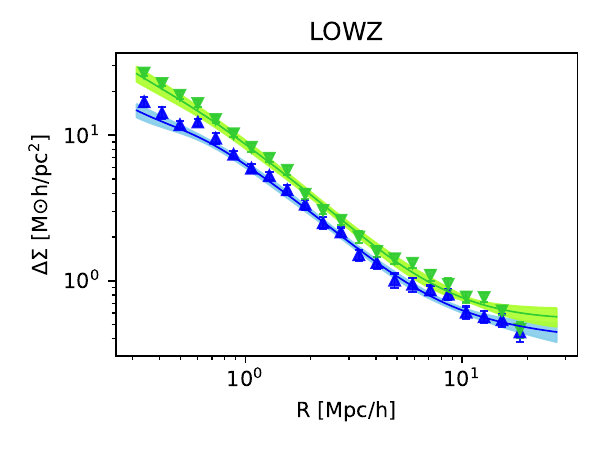}
    \caption{The measured stacked weak lensing signal for bins of catalogs, overlaid with the best fitting model. In the first row, the signal from the redMaPPer, ZOU21, and YANG21 catalogs are shown from left to right. In these three panels, the signals from the low-z low-M, low-z high-M, high-z low-M, and high-z high-M are shown in blue, green, red, and grey, respectively.
    In the second row, the signals from CMASS and LOWZ catalogs are shown in the left panel and the right panel. In these two panels, the low-M and high-M bins are shown in blue and green, respectively. 
    For each bin, the best-fitting model is shown in the same color, together with a lighter shade region for its 1$\sigma$ error range.}
    \label{fig:ggl_signal}
\end{figure*}

The projected mass density of the lens, $\Sigma$, is known related to the azimuthally averaged tangential shear at projected radius $R$. However, the differential surface density, $\Delta\Sigma(R)$, is usually used for the measurement of weak lensing signal, which is defined as 
\begin{equation}
    \Delta\Sigma=\bar\Sigma(<R)-\Sigma(R).
\end{equation}
We use the $SWOT$\footnote{http://jeancoupon.com/swot} software \citep{Coupon2012} to measure the weak lensing signal for every halo bin, as shown in Fig.~\ref{fig:ggl_signal}. The parameter setting is listed in Tab.~\ref{tab:swot_para}.

\begin{table}[t]
    \centering
    \footnotesize
    \caption{Setting of $SWOT$ parameters}
    \begin{tabular}{c|c|l}
    \hline
    \hline
    Par. & Value & Meaning \\
    \hline
        corr                & gglens        & Type of correlation\\
        range               & 0.1-30~Mpc/h  & Correlation range\\
        nbins               & 30            & Number of bins \\
        err                 & Jackknife     & Resampling method\\
        nsub                & 64	        & Number of resampling subvolumes \\
        H$_0$               & 67.4          & Hubble parameter\\
        $\Omega_{\rm m}$    & 0.315         & Relative matter density\\
        $\Omega_{\rm L}$    & 0.684         & Relative energy density\\
        $\Delta$            & 0.1           & Minimum redshift difference \\
                            &               & between the source and the lens\\
        proj                & como          & Projection\\
    \hline
    \hline
    \end{tabular}\newline
    {{\bf Note.} The parameter name, value, and corresponding physical meaning are presented in columns.}
    \label{tab:swot_para}
\end{table}

The differential surface density is estimated as
\begin{equation}
\Delta\Sigma(R)=\frac{\sum_{\rm ls}w_{\rm ls}~\gamma_t^{\rm ls}~\Sigma_{\rm crit}}{\sum_{\rm ls}w_{\rm ls}}\,,
\end{equation}
where $\gamma_t^{\rm ls}$ is the tangential shear, $w_{\rm ls}=w_{\rm n}\Sigma_{\rm crit}^{-2}$, and $w_{\rm n}$ is the weight factor to account for intrinsic scatter in ellipticity and the error of shape measurement \citep{Miller2007, Miller2013}. The $w_{\rm n}$ used in this work is defined as $w_{\rm n}=1/(\sigma^2_{\epsilon}+\sigma^2_{\rm e})$. The $\sigma_{\epsilon}$ is the intrinsic ellipticity dispersion derived from the whole galaxy sample, and taken as $0.27$ \citep{Giblin2021}. $\sigma_{\rm e}$ is the error of the ellipticity measurement \citep{Hoekstra2002}.

The lensing signal is recalibrated as
\begin{equation}
\Delta\Sigma^{\rm cal}(R)=\frac{\Delta\Sigma(R)}{1+K(z_{\rm l})},
\end{equation}
and \begin{equation}
1+K(z_{\rm l})=\frac{\sum_{\rm ls}w_{\rm ls}~(1+m)}{\sum_{\rm ls}w_{\rm ls}},
\end{equation}
where $m$ is the multiplicative error.

In the measurement of galaxy-galaxy lensing signal, there are multiple potential systematics. 
Potential uncertainties may arise due to the shear additive and multiplicative bias, the intrinsic alignments, the boost factor, and photo-$z$ dilution.
Firstly, the shear additive bias, which is expected from the residual of the shape fitting from anisotropic point spread function correction \citep{Zuntz2018, McClintock2019}, brings only negligible bias to our measurement of galaxy-galaxy lensing signal.
We cross-match DECaLS sources with external
shear measurements \citep{Phriksee2020, Yao2020, Zu2021}, including Canada-France-Hawaii 
Telescope Stripe 82 \citep{Moraes2014}, Dark Energy Survey
\citep{DES2016}, and Kilo-Degree Survey \citep{Hildebrandt2017} sources. This way, we obtain the multiplicative and additive biases. After correction, we expect the residual multiplicative bias of DECaLS DR8 shear catalog is $m\sim5\%$ \citep{Yao2020, Phriksee2020}, possibly from the selection of observation data \citep{Li2021, Jarvis2016} (i.e. the matched galaxies are not an exact representation of the full observational sample in terms of magnitude, color, size, etc.), or the slight differences between simulated galaxies and observations.

Furthermore, we only use bright galaxies with $r$-band magnitude $<23$ as sources \citep{Zou2019} to avoid photo-$z$ outliers and keep the overlap between lens and sources at a very small level. In this case, both effects from intrinsic alignment \citep{Yao2023, Yao2023b}, and the one from boost factor are negligible \citep{Melchior2017, McClintock2019}.
For the photo-$z$ dilution \citep{Lange2021}, we have not made the n($z$) correction. Nevertheless, this is a scale-independent problem, similar to the residual shear multiplicative bias, which does not affect our splashback radius measurement.

\subsection{Fitting of galaxy-galaxy lensing signal}

In our measurement, the contribution of $\Delta\Sigma$ includes the central dark matter halo, the miscentering effect \citep{Johnston2007}, and nearby halos.
To avoid the central region where the contribution from the central galaxy dominates, we only consider the signal within the radius range of $0.3-30$~Mpc/$h$ in the fitting.
Thus, the whole model used is 
\begin{equation}
\begin{split}
\Delta\Sigma(R)=(1-f_{\rm mis})~\Delta\Sigma_{\rm ein}(R)~~~~~~~~~~~~~~~~~~~~~\\
+ f_{\rm mis} ~\Delta\Sigma_{\rm mis}(R|R_{\rm mis})
+\Delta\Sigma_{\rm outer}(R).
\end{split}
\label{eq:deltasigma}
\end{equation}

We use the Einasto profile \citep{Einasto1965} for the contribution of the central dark matter halo.
Compared with the Navarro-Frenk-White profile (NFW profile, \citealt{Navarro1996}), the Einasto profile can characterize the dark matter halo profile more accurately \citep{Navarro2004, Navarro2010, Gao2008, Ludlow2011}. 

The miscentering effect comes from the inaccurate determination of the halo center \citep{Johnston2007}, which greatly reduces the stacked signal at the central region. 
For a cluster offset from the cluster center by the distance of $R_{\rm mis}$, its surface mass density becomes $\Sigma_{\rm mis}=\int^{2\pi}_0 {{\rm d}\theta}~\Sigma(\sqrt{R^2+R_{\rm mis}^2+2RR_{\rm mis} \rm{cos} \theta})/{2\pi}$. The Gamma
profile is assumed to obtain the miscentering signal of the stacked signal \citep{McClintock2019}. $f_{\rm mis}$ and $R_{\rm mis}$ are the two characteristic parameters, for the fraction of offset halos and the offset distance, respectively. 

The last term in Eq.~\ref{eq:deltasigma} is the outer term indicating the signal from nearby halos, which dominates at the cluster outskirt. This term is estimated from the non-linear 
scaling of the matter power spectrum as a function of redshift with the $Halofit$ model using the $CAMB$ package\footnote{https://github.com/cmbant/CAMB} \citep{Lewis2002, Lewis1999, Howlett2012}.

\subsection{Estimation of splashback radius}

For estimation of the splashback radius, we need to obtain the profile of the halo density. In this work, we use the density profile from \cite{Diemer2014} (DK14 profile for short):
\begin{equation}
\begin{split}
\rho(r)=\rho_{\rm inner}\times f_{\rm trans} + \rho_{\rm outer}, ~~~~~~~~~~~~~~~~\\
\rho_{\rm inner}= \rho_{\rm Einasto} = \rho_{\rm s}~ {\rm exp}(-\frac{2}{\alpha} [(\frac{r}{r_{\rm s}})^{\alpha}-1]), \\
f_{\rm trans}=[1+(\frac{r}{r_{\rm t}})^\beta]^{-\frac{\gamma}{\beta}},~~~~~~~~~~~~~~~~~~~~~~~~~\\
\rho_{\rm outer}=\frac{\rm Norm}{(\frac{r}{5R_{\rm 200m}})^{\rm slope} +1}.~~~~~~~~~~~~~~~~~~~~~
\end{split}
\label{eq:rho}
\end{equation}

The DK14 profile exhibits an Einasto profile in its inner region, transitioning to a steeper gradient near the virial radius. The Einasto profile incorporates six free parameters: 
$\rho_{\rm s}$, representing the density at the scale radius; 
$r_{\rm s}$, the scale radius; 
$r_{\rm t}$, the truncation radius where the profile steepens beyond the Einasto profile; 
$\alpha$, a factor determining the rate at which the slope of the inner Einasto profile increases; 
$\beta$, the sharpness of the steepening; 
$\gamma$, the asymptotic negative incline of the steepening term. 
Besides, the normalization and slope from the outer term are also free parameters. 
Furthermore, the distance and fraction of the miscentering are other free parameters.
We use the $Colossus$ software\footnote{https://bdiemer.bitbucket.io/colossus/} \citep{Diemer2018}, and the $Cluster\_toolkit$ software\footnote{https://cluster-toolkit.readthedocs.io/en/latest/index.html} \citep{McClintock2019} to model the density profile and the corresponding surface density profile. 

We make the model-fitting using the Markov Chain Monte Carlo (MCMC) technique ($EMCEE$ package\footnote{https://emcee.readthedocs.io/en/stable/}, \citealt{Foreman-Mackey2013}). The number of chains is $50$, with $6,000$ steps each. The priors are set according to Tab.~\ref{tab:prior}, with the starting value of each chain exhibiting a variation of $0.1\%$ around the specified value. To minimize the bias originating from priors, we discard the initial $3,000$ steps. One parameter set is retained for every three sets. 

\begin{table}[t]
\small
\centering
\caption{Prior ranges of parameters.} 
\begin{tabular}{c|c|c|c}
\hline
\hline
Par.                              & Prior value       & Min.        & Max.       \\
\hline
$\rho_{\rm s}$ [$10^{12}$M$_\odot$ h$^2$/Mpc$^3$]& 10            & 0.1        & 1,000 \\ 
$r_{\rm s}$ [Mpc/h]                     & 0.40               & 0.10        & 1.50      \\
$r_{\rm t}$ [Mpc/h]                     & 1.84             & 1.00      & 6.00      \\
log($\alpha$)                           & [log(0.22), 0.6]  & log(0.01)  & log(10)   \\
log($\beta$)                            & [log(6), 0.2]     & log(0.01)  & log(30)   \\
log($\gamma$)                           & [log(4), 0.2]     & log(0.01)  & log(30)   \\
Norm                                    & 300               & 50         & 3,000      \\
slope                                   & 0.9               & 0.1        & 0.99      \\
$R_{\rm mis}$ [Mpc/h]                   & 0.55              & 0.0        & 1.0       \\
$f_{\rm mis}$                           & 0.2               & 0.0        & 1.0       \\
\hline
\hline
\end{tabular}\newline
{{\bf Note.} The prior value, minimum, and maximum values of parameters are presented in columns. The $\alpha$, $\beta$, and $\gamma$ are logarithm Gaussian distributed, with the mean and full width at half maximum (FWHM) shown in the $Prior~value$ column. The prior settings for $\alpha$, $\beta$, and $\gamma$ are based on \citet{Shin2019, Shin2021, Adhikari2021, Dacunha2022}.}
\label{tab:prior}
\end{table}

By deriving the best-fit density profile for the given cluster sample, we obtain the MCMC chain which describes the posterior distribution of our model parameters for the density profile. Then we calculate the $d\log(\rho)/d\log(r)$ for each point within the MCMC chain. Utilizing the equation sets outlined in Eq.~\ref{eq:rho}, we derive $R_{\rm sp}$ as the radius with the minimum value of the density slope, $d\log{\rho(r)}/d\log{r}$.

This way, we obtain the posterior distribution of $R_{\rm sp}$. Subsequently, we determine the best fit $R_{\rm sp}$ by averaging the posterior parameter space, and its associated uncertainty is given by calculating the confidence level 68$\%$ of the posterior distribution.

The modeling of the infalling term does not affect the best-fit $R_{\rm sp}$ in our measurement. In the model, the contribution from the infalling material is modeled as the power-law term. And the splashback radius is still notable in the profile of the density slope. 

\section{Result and discussion}
\label{sec:result}

\subsection{Measurement of the splashback radius}

\begin{figure*}
    \centering
    \includegraphics[width=0.7\textwidth]{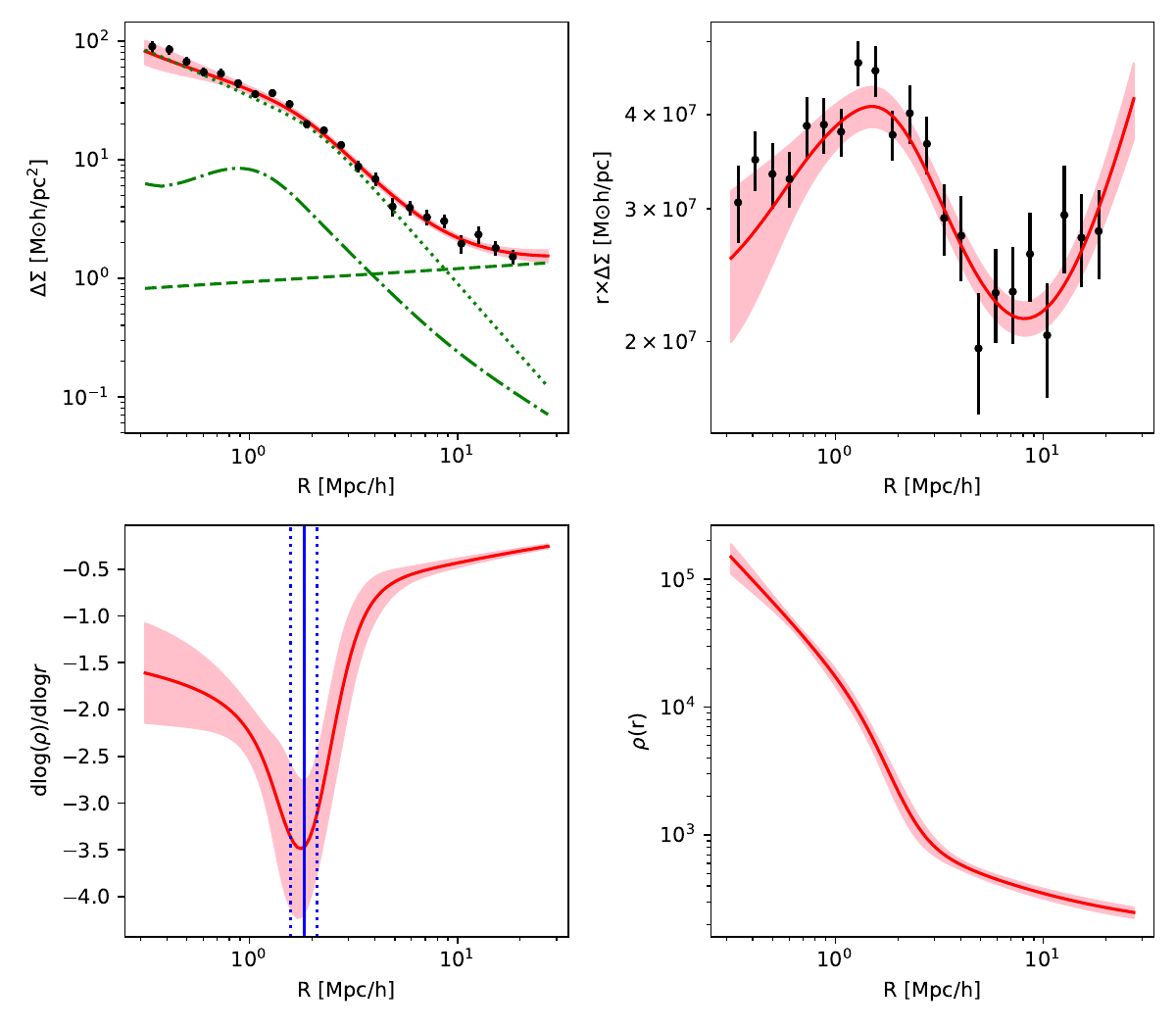}\\
    \caption{The panels show the profile of the differential surface density ($\Delta\Sigma$(R)), $r\times\Delta\Sigma$(R), the density slope, and the density for the fitting result of redMaPPer halos in the low-redshift high-mass bins. The data points with error bars are represented by black dots, while the optimal model and its 1$\sigma$ range are represented by a red curve and pink shading, respectively. In the upper-left panel, the contributions from the central halo, miscentering effect, and nearby halo are overlaid as dotted, dot-dashed, and dashed green curves. In the lower-left panel, the best-fit splashback radius and its 1$\sigma$ range are displayed as the solid blue vertical line and dotted blue vertical lines.}
    \label{fig:result_example}
\end{figure*}

The fitting results for the low redshift, high mass bin of redMaPPer clusters are illustrated in Fig.~\ref{fig:result_example} as an example. 
The figure comprises four panels, presenting the galaxy-galaxy lensing signal with the best-fitting model and its model components, the differential surface density multiplied by the radius, the density slope, and the density profile. 
In the lower-left panel, for each set of parameters in the MCMC chains, a profile of the density slope is obtained, and the best-fit model and its 1$\sigma$ range are estimated at each radius. On the other hand, the best-fitting splashback radius and its 1$\sigma$ range are calculated as the minimum location of each profile. As a result, the value of splashback radius and its $1\sigma$ range does not necessarily match with the bottom range of the distribution of the $d\log(\rho)/d\log(r)$.

The values of the optimal fitting parameters are listed on Tab.~\ref{tab:fit_para}. Notably, the reasonable reduced-$\chi^2$ values of the best fitting in the last column indicate that our model effectively characterizes the features of the galaxy-galaxy lensing signal.

\begin{table*}
 \begin{center}
  \caption{The optimal parameter fitting results.}
    \footnotesize
       \begin{tabular}{c c c c c c c c c c c c c }
        \hline
        \hline
   Cat. & bin &    $\rho_s$ & $r_{\rm s}$& $r_{\rm t}$ & log($\alpha$)& log($\beta$)& log($\gamma$) & Norm & Slope & $R_{\rm mis}$ & $f_{\rm mis}$ &$\chi^2/\nu$ \\
(1) & (2) & (3) & (4) & (5) & (6) & (7) & (8) & (9) & (10) & (11) & (12)& (13)    \\
       \hline
RM &z1m1 &14.20$^{+12.71}_{-8.21}$ &0.62$^{+0.35}_{-0.20}$ &3.68$^{+1.57}_{-1.48}$ &-0.89$^{+0.41}_{-0.50}$ &0.77$^{+0.20}_{-0.22}$ &0.57$^{+0.19}_{-0.19}$ &342$^{+214}_{-188}$ &0.89$^{+0.07}_{-0.11}$ &0.33$^{+0.02}_{-0.02}$ &0.57$^{+0.18}_{-0.22}$ &1.60 \\
RM &z1m2 &33.54$^{+23.29}_{-13.08}$ &0.58$^{+0.20}_{-0.15}$ &1.47$^{+1.07}_{-0.34}$ &-1.08$^{+0.97}_{-0.39}$ &0.73$^{+0.22}_{-0.24}$ &0.67$^{+0.20}_{-0.18}$ &712$^{+427}_{-452}$ &0.89$^{+0.07}_{-0.15}$ &0.46$^{+0.30}_{-0.19}$ &0.16$^{+0.13}_{-0.10}$ &2.78 \\
RM &z2m1 &17.57$^{+26.07}_{-10.76}$ &0.55$^{+0.35}_{-0.21}$ &2.75$^{+1.70}_{-1.28}$ &-0.97$^{+0.56}_{-0.54}$ &0.74$^{+0.22}_{-0.23}$ &0.59$^{+0.21}_{-0.20}$ &141$^{+126}_{-65}$ &0.91$^{+0.06}_{-0.10}$ &0.31$^{+0.21}_{-0.05}$ &0.26$^{+0.21}_{-0.18}$ &1.56 \\
RM &z2m2 &40.03$^{+57.44}_{-29.87}$ &0.42$^{+0.46}_{-0.15}$ &3.08$^{+1.94}_{-1.55}$ &-0.72$^{+0.40}_{-0.60}$ &0.74$^{+0.22}_{-0.20}$ &0.62$^{+0.19}_{-0.19}$ &211$^{+199}_{-113}$ &0.91$^{+0.06}_{-0.12}$ &0.39$^{+0.09}_{-0.05}$ &0.50$^{+0.26}_{-0.30}$ &1.28 \\
ZOU21 &z1m1 &10.20$^{+12.77}_{-5.21}$ &0.45$^{+0.20}_{-0.16}$ &5.23$^{+0.55}_{-0.91}$ &-1.43$^{+0.30}_{-0.31}$ &0.90$^{+0.16}_{-0.17}$ &0.57$^{+0.20}_{-0.19}$ &221$^{+108}_{-95}$ &0.92$^{+0.05}_{-0.09}$ &0.31$^{+0.02}_{-0.02}$ &0.38$^{+0.09}_{-0.09}$ &4.19 \\
ZOU21 &z1m2 &11.13$^{+7.37}_{-4.51}$ &0.62$^{+0.21}_{-0.15}$ &5.10$^{+0.62}_{-0.92}$ &-0.89$^{+0.19}_{-0.24}$ &0.84$^{+0.19}_{-0.20}$ &0.55$^{+0.19}_{-0.20}$ &321$^{+127}_{-143}$ &0.93$^{+0.05}_{-0.09}$ &0.36$^{+0.02}_{-0.02}$ &0.52$^{+0.07}_{-0.10}$ &2.26 \\
ZOU21 &z2m1 &25.35$^{+24.34}_{-15.51}$ &0.28$^{+0.17}_{-0.08}$ &4.85$^{+0.82}_{-1.12}$ &-1.12$^{+0.33}_{-0.39}$ &0.82$^{+0.19}_{-0.21}$ &0.59$^{+0.18}_{-0.18}$ &139$^{+97}_{-65}$ &0.92$^{+0.05}_{-0.09}$ &0.34$^{+0.29}_{-0.05}$ &0.21$^{+0.26}_{-0.16}$ &2.09 \\
ZOU21 &z2m2 &12.50$^{+10.73}_{-6.77}$ &0.53$^{+0.27}_{-0.15}$ &4.95$^{+0.77}_{-1.33}$ &-0.99$^{+0.34}_{-0.42}$ &0.84$^{+0.18}_{-0.23}$ &0.55$^{+0.19}_{-0.19}$ &204$^{+118}_{-98}$ &0.93$^{+0.05}_{-0.10}$ &0.37$^{+0.03}_{-0.03}$ &0.54$^{+0.11}_{-0.19}$ &2.39 \\
YANG21 &z1m1 &12.12$^{+10.13}_{-6.53}$ &0.35$^{+0.15}_{-0.09}$ &2.94$^{+2.07}_{-1.65}$ &-0.76$^{+0.46}_{-0.49}$ &0.75$^{+0.23}_{-0.29}$ &0.59$^{+0.20}_{-0.21}$ &151$^{+145}_{-77}$ &0.92$^{+0.05}_{-0.09}$ &0.25$^{+0.06}_{-0.04}$ &0.41$^{+0.29}_{-0.27}$ &3.08 \\
YANG21 &z1m2 &12.89$^{+9.79}_{-6.49}$ &0.42$^{+0.18}_{-0.12}$ &3.50$^{+1.76}_{-2.15}$ &-0.92$^{+0.32}_{-0.37}$ &0.72$^{+0.23}_{-0.35}$ &0.60$^{+0.21}_{-0.22}$ &252$^{+145}_{-120}$ &0.92$^{+0.05}_{-0.10}$ &0.29$^{+0.40}_{-0.02}$ &0.41$^{+0.29}_{-0.28}$ &4.01 \\
YANG21 &z2m1 &15.06$^{+29.91}_{-9.74}$ &0.28$^{+0.20}_{-0.11}$ &4.58$^{+0.97}_{-1.68}$ &-1.18$^{+0.46}_{-0.42}$ &0.76$^{+0.19}_{-0.18}$ &0.61$^{+0.19}_{-0.22}$ &149$^{+97}_{-67}$ &0.93$^{+0.05}_{-0.08}$ &0.22$^{+0.15}_{-0.10}$ &0.22$^{+0.36}_{-0.17}$ &1.27 \\
YANG21 &z2m2 &11.91$^{+14.80}_{-7.04}$ &0.39$^{+0.24}_{-0.12}$ &4.63$^{+0.97}_{-1.76}$ &-1.02$^{+0.44}_{-0.47}$ &0.80$^{+0.20}_{-0.21}$ &0.56$^{+0.20}_{-0.18}$ &161$^{+104}_{-75}$ &0.92$^{+0.05}_{-0.10}$ &0.28$^{+0.03}_{-0.03}$ &0.43$^{+0.32}_{-0.26}$ &1.83 \\
CMASS &m1 &31.11$^{+24.23}_{-19.41}$ &0.16$^{+0.11}_{-0.04}$ &3.89$^{+1.37}_{-1.99}$ &-0.97$^{+0.28}_{-0.23}$ &0.76$^{+0.21}_{-0.23}$ &0.64$^{+0.24}_{-0.22}$ &95$^{+52}_{-27}$ &0.94$^{+0.04}_{-0.06}$ &0.32$^{+0.51}_{-0.21}$ &0.08$^{+0.12}_{-0.06}$ &1.34 \\
CMASS &m2 &38.44$^{+35.19}_{-23.48}$ &0.20$^{+0.10}_{-0.04}$ &3.53$^{+1.61}_{-2.13}$ &-0.53$^{+0.23}_{-0.30}$ &0.75$^{+0.20}_{-0.25}$ &0.60$^{+0.21}_{-0.20}$ &122$^{+52}_{-55}$ &0.94$^{+0.04}_{-0.08}$ &0.26$^{+0.44}_{-0.06}$ &0.15$^{+0.15}_{-0.12}$ &3.30 \\
LOWZ &m1 &13.39$^{+9.46}_{-6.38}$ &0.30$^{+0.12}_{-0.06}$ &3.44$^{+1.75}_{-1.99}$ &-0.71$^{+0.29}_{-0.40}$ &0.73$^{+0.23}_{-0.21}$ &0.59$^{+0.19}_{-0.21}$ &232$^{+134}_{-116}$ &0.93$^{+0.04}_{-0.06}$ &0.25$^{+0.04}_{-0.03}$ &0.27$^{+0.22}_{-0.19}$ &1.66 \\
LOWZ &m2 &15.97$^{+8.70}_{-5.52}$ &0.39$^{+0.09}_{-0.09}$ &2.05$^{+1.83}_{-0.81}$ &-1.38$^{+0.41}_{-0.31}$ &0.21$^{+0.36}_{-0.17}$ &0.46$^{+0.25}_{-0.22}$ &397$^{+119}_{-167}$ &0.95$^{+0.03}_{-0.07}$ &0.41$^{+0.53}_{-0.19}$ &0.06$^{+0.14}_{-0.05}$ &4.56 \\
    \hline
    \hline
        \end{tabular}
        \footnotesize
     {Note: The first two columns show the catalog name and bin names, while the subsequent ten columns present the parameters for the fitting of differential surface density profile, $\Delta\Sigma$(R), in Eq.~\ref{eq:deltasigma} and Eq.~\ref{eq:rho}. The last column denotes the reduced $\chi^2$ for the best fit. 
    The unit of $\rho_s$ is $10^{12}$M$_\odot$h$^2$/Mpc$^3$. The $r_{\rm s}$, $r_{\rm t}$, and $R_{\rm mis}$ are all in the unit of Mpc/$h$. All listed errors correspond to the 1$\sigma$ range.}
    \label{tab:fit_para}
 \end{center}
\end{table*}

In addition to our measurements, we have compiled a list of splashback radius measurements from the literature. These are presented in the first part of Tab.~\ref{tab:sbr_previous}, with our results in the second part for comparison. 
The sample listed in Tab.~\ref{tab:sbr_previous}, including both literature results and our measurements, is named as ``the whole sample" in later text.
These literature works make measurements mainly using weak lensing signal and galaxy number density profiles, for various cluster catalogs, including DES redMaPPer \citep{Chang2018, Shin2019}, HSC redMaPPer \citep{Murata2020}, X-ray clusters \citep{Contigiani2019, Bianconi2021}, SZ-selected clusters \citep{Shin2019, Zurcher2019, Adhikari2021, Shin2021}, and luminous red galaxies from KiDS \citep{Contigiani2023}. 

In compiling literature works, we have standardized the values to the same unit and accounted for differences in the Hubble constant. When both the values of $R_{\rm sp}$ and $R_{\rm sp}$/$R_{\rm 200m}$ are provided, we calculate the $R_{\rm 200m}$ values from them; otherwise, the $R_{\rm 200m}$ value is calculated from $M_{\rm 200m}$ and the redshift. The corresponding errors are propagated as Eq.~\ref{eq:error_propagate}. The concentration value listed in Tab.~\ref{tab:sbr_previous} is one of the fitting parameters in our model. It is converted to the value at $z=0$ by multiplying $(1+z)^{0.67}$ \citep{Klypin2016}. 

\begin{equation}
\begin{split}
(\sigma_z/z)=a*(\sigma_x/x), ~~~~~~~~~~~~~~~(z=x^a)\\
~~~~~(\sigma_z/z)^2=(\sigma_x/x)^2+(\sigma_y/y)^2, ~~~(z=x*y)\\
~~~~(\sigma_z/z)^2=(\sigma_x/x)^2+(\sigma_y/y)^2, ~~~~(z=x/y)
\end{split}
\label{eq:error_propagate}
\end{equation}

\begin{table*}[t]
   \centering
\caption{Splashback radius measurements in the ``whole sample", including literature and our work.}
   \small
    \begin{tabular}{c|c|c|c|c|c|c|c|c}
    \hline
    \hline     
    Index &Mean $M_{\rm 200m}$ & $z$ & $R_{\rm sp}$[Mpc/h]    &   $R_{\rm 200m}$[Mpc/h] &  $R_{\rm sp}$/$R_{\rm 200m}\/$      & $c(z=0)$  & Sample            &  Measurement               \\
    (1) &(2) &(3) &(4) &(5) &(6) &(7) &(8) &(9) \\
    \hline
     1& ${  1.69}^{+0.31}_{-0.31}$ & ${0.41}$ & ${1.13}^{+0.07}_{-0.07}$ & ${0.78}^{+0.05}_{-0.05}$ & ${0.82}^{+0.05}_{-0.05}$ & - &               DES RM&           G profile\\             
     2& ${  1.69}^{+0.79}_{-0.79}$ & ${0.41}$ & ${1.34}^{+0.21}_{-0.21}$ & ${0.78}^{+0.12}_{-0.12}$ & ${0.97}^{+0.15}_{-0.15}$ & - &               DES RM&                  WL\\             
     3& ${  1.71}^{+0.08}_{-0.07}$ & ${0.57}$ & ${1.50}^{+0.19}_{-0.18}$ & ${1.32}^{+0.02}_{-0.02}$ & ${1.14}^{+0.14}_{-0.14}$ & - &               HSC RM&           G profile\\            
     4& ${ 11.46}^{+10.78}_{-6.85}$ & ${0.28}$ & ${2.36}^{+0.74}_{-0.47}$ & ${1.54}^{+0.48}_{-0.31}$ & ${1.53}^{+0.68}_{-0.43}$ & - &           CCCP X-ray&                  WL\\       
     5& ${ 14.03}^{+0.12}_{-0.12}$ & ${0.23}$ & ${3.81}^{+0.75}_{-0.75}$ & ${2.18}^{+0.01}_{-0.01}$ & ${1.74}^{+0.34}_{-0.34}$ & - &         LoCuss X-ray&           G profile\\          
     6& ${  5.10}^{+1.56}_{-1.34}$ & ${0.46}$ & ${1.06}^{+0.11}_{-0.09}$ & ${1.10}^{+0.11}_{-0.10}$ & ${0.97}^{+0.07}_{-0.06}$ & - &               DES RM&           G profile\\              
     7& ${  5.10}^{+4.61}_{-4.43}$ & ${0.49}$ & ${1.32}^{+0.40}_{-0.38}$ & ${1.08}^{+0.33}_{-0.31}$ & ${1.22}^{+0.26}_{-0.25}$ & - &               SPT SZ&           G profile\\              
     8& ${  5.60}^{+7.71}_{-5.99}$ & ${0.49}$ & ${1.24}^{+0.57}_{-0.44}$ & ${1.12}^{+0.51}_{-0.40}$ & ${1.11}^{+0.36}_{-0.28}$ & - &               ACT SZ&           G profile\\              
     9& ${  5.87}^{+2.47}_{-2.87}$ & ${0.18}$ & ${1.78}^{+0.25}_{-0.29}$ & ${1.28}^{+0.18}_{-0.21}$ & ${1.39}^{+0.28}_{-0.32}$ & - &            Planck SZ&           G profile\\           
    10& ${  4.62}^{+3.50}_{-4.90}$ & ${0.46}$ & ${1.19}^{+0.30}_{-0.42}$ & ${1.06}^{+0.27}_{-0.38}$ & ${1.12}^{+0.20}_{-0.28}$ & - &               ACT SZ&                  WL\\              
    11& ${  4.62}^{+1.11}_{-2.40}$ & ${0.46}$ & ${1.13}^{+0.09}_{-0.20}$ & ${1.06}^{+0.09}_{-0.18}$ & ${1.06}^{+0.06}_{-0.13}$ & - &               ACT SZ&           G profile\\              
    12& ${  5.77}^{+2.17}_{-2.90}$ & ${0.49}$ & ${2.39}^{+0.30}_{-0.40}$ & ${1.13}^{+0.14}_{-0.19}$ & ${2.11}^{+0.38}_{-0.50}$ & - &           ACT DR5 SZ&          G profiles\\          
    13& ${  0.38}^{+0.24}_{-0.14}$ & ${0.44}$ & ${1.00}^{+0.13}_{-0.13}$ & ${0.47}^{+0.06}_{-0.06}$ & ${2.15}^{+0.39}_{-0.39}$ & - &                 KiDS&          Splash-back\\       
    \hline
    14& ${  3.38}^{+0.82}_{-0.82}$ & ${0.2620}^{+0.0671}_{-0.0670}$ & ${1.96}^{+1.05}_{-1.05}$ & ${1.03}^{+0.08}_{-0.08}$ & ${1.90}^{+1.03}_{-1.03}$ & ${4.95}^{+1.62}_{-1.62}$ &              RM,z1m1&                  WL\\           
    15& ${  9.10}^{+2.54}_{-2.54}$ & ${0.2564}^{+0.0682}_{-0.0682}$ & ${1.84}^{+0.27}_{-0.27}$ & ${1.44}^{+0.13}_{-0.13}$ & ${1.27}^{+0.22}_{-0.22}$ & ${7.46}^{+2.00}_{-2.00}$ &              RM,z1m2&                  WL\\           
    16& ${  3.10}^{+1.06}_{-1.06}$ & ${0.3962}^{+0.0317}_{-0.0317}$ & ${2.04}^{+0.79}_{-0.79}$ & ${0.95}^{+0.11}_{-0.11}$ & ${2.14}^{+0.86}_{-0.86}$ & ${5.83}^{+2.55}_{-2.55}$ &              RM,z2m1&                  WL\\           
    17& ${  4.22}^{+1.18}_{-1.18}$ & ${0.4472}^{+0.0528}_{-0.0527}$ & ${2.06}^{+0.85}_{-0.85}$ & ${1.04}^{+0.10}_{-0.10}$ & ${1.98}^{+0.84}_{-0.84}$ & ${7.71}^{+3.40}_{-3.39}$ &              RM,z2m2&                  WL\\           
    18& ${  0.86}^{+0.08}_{-0.08}$ & ${0.3403}^{+0.0585}_{-0.0585}$ & ${0.44}^{+0.17}_{-0.17}$ & ${0.63}^{+0.02}_{-0.02}$ & ${0.69}^{+0.27}_{-0.27}$ & ${4.63}^{+1.57}_{-1.56}$ &           ZOU21,z1m1&                  WL\\           
    19& ${  2.52}^{+0.35}_{-0.35}$ & ${0.3459}^{+0.0579}_{-0.0578}$ & ${1.22}^{+0.42}_{-0.42}$ & ${0.91}^{+0.04}_{-0.04}$ & ${1.35}^{+0.47}_{-0.47}$ & ${4.49}^{+1.05}_{-1.05}$ &           ZOU21,z1m2&                  WL\\           
    20& ${  0.71}^{+0.15}_{-0.15}$ & ${0.5040}^{+0.0428}_{-0.0429}$ & ${0.60}^{+0.25}_{-0.25}$ & ${0.56}^{+0.04}_{-0.04}$ & ${1.08}^{+0.45}_{-0.45}$ & ${6.52}^{+2.27}_{-2.27}$ &           ZOU21,z2m1&                  WL\\           
    21& ${  1.62}^{+0.21}_{-0.21}$ & ${0.5074}^{+0.0433}_{-0.0434}$ & ${0.97}^{+0.45}_{-0.45}$ & ${0.74}^{+0.03}_{-0.03}$ & ${1.32}^{+0.61}_{-0.61}$ & ${4.65}^{+1.37}_{-1.37}$ &           ZOU21,z2m2&                  WL\\           
    22& ${  0.49}^{+0.34}_{-0.34}$ & ${0.3365}^{+0.0467}_{-0.0467}$ & ${0.94}^{+0.36}_{-0.36}$ & ${0.53}^{+0.12}_{-0.12}$ & ${1.78}^{+0.79}_{-0.79}$ & ${4.51}^{+1.36}_{-1.36}$ &          YANG21,z1m1&                  WL\\           
    23& ${  1.00}^{+0.59}_{-0.59}$ & ${0.3367}^{+0.0469}_{-0.0469}$ & ${0.94}^{+0.40}_{-0.40}$ & ${0.67}^{+0.13}_{-0.13}$ & ${1.41}^{+0.66}_{-0.66}$ & ${4.71}^{+1.31}_{-1.31}$ &          YANG21,z1m2&                  WL\\           
    24& ${  0.37}^{+0.16}_{-0.16}$ & ${0.4694}^{+0.0348}_{-0.0348}$ & ${0.51}^{+0.26}_{-0.26}$ & ${0.45}^{+0.06}_{-0.06}$ & ${1.11}^{+0.59}_{-0.59}$ & ${5.58}^{+2.61}_{-2.60}$ &          YANG21,z2m1&                  WL\\           
    25& ${  0.60}^{+0.19}_{-0.19}$ & ${0.4707}^{+0.0353}_{-0.0353}$ & ${0.70}^{+0.39}_{-0.39}$ & ${0.54}^{+0.06}_{-0.06}$ & ${1.31}^{+0.74}_{-0.74}$ & ${4.64}^{+1.75}_{-1.75}$ &          YANG21,z2m2&                  WL\\           
    26& ${  0.19}^{+0.04}_{-0.04}$ & ${0.5219}^{+0.0563}_{-0.0563}$ & ${0.58}^{+0.32}_{-0.32}$ & ${0.35}^{+0.03}_{-0.03}$ & ${1.62}^{+0.92}_{-0.92}$ & ${6.78}^{+2.36}_{-2.36}$ &             CMASS,m1&                  WL\\           
    27& ${  0.36}^{+0.06}_{-0.06}$ & ${0.5467}^{+0.0565}_{-0.0566}$ & ${0.89}^{+0.25}_{-0.25}$ & ${0.44}^{+0.03}_{-0.03}$ & ${2.04}^{+0.58}_{-0.58}$ & ${7.21}^{+2.19}_{-2.19}$ &             CMASS,m2&                  WL\\           
    28& ${  0.39}^{+0.14}_{-0.14}$ & ${0.2379}^{+0.0726}_{-0.0727}$ & ${0.88}^{+0.30}_{-0.30}$ & ${0.51}^{+0.06}_{-0.06}$ & ${1.72}^{+0.62}_{-0.62}$ & ${4.77}^{+1.22}_{-1.22}$ &              LOWZ,m1&                  WL\\           
    29& ${  1.17}^{+0.47}_{-0.47}$ & ${0.2691}^{+0.0565}_{-0.0564}$ & ${0.79}^{+0.22}_{-0.22}$ & ${0.72}^{+0.10}_{-0.10}$ & ${1.10}^{+0.34}_{-0.34}$ & ${5.44}^{+1.05}_{-1.05}$ &              LOWZ,m2&                  WL\\           
    \hline
    \hline
    \end{tabular}
    The column is, in sequence, the index, halo mass, redshift, splashback radius, $R_{\rm 200m}$, normalized splashback radius, concentration at $z=0$, sample name, and measurement method. In the last column, WL is short for weak lensing method. 
    The first 13 rows are obtained from literature, whose reference is as follows. 1, \cite{Chang2018}; 2, \cite{Chang2018}; 3, \cite{Murata2020}; 4, \cite{Contigiani2019}; 5, \cite{Bianconi2021}; 6, \cite{Shin2019}; 7, \cite{Shin2019}; 8, \cite{Shin2019}; 9, \cite{Zurcher2019}; 10, \cite{Shin2021}; 11, \cite{Shin2021}; 12, \cite{Adhikari2021}; 13, \cite{Contigiani2023}. 
    The data listed in the 14th to 29th rows are the results of this work.
    \label{tab:sbr_previous}
\end{table*}

\subsection{Dependence of $R_{\rm sp}$ on halo mass, redshift, and $R_{\rm 200m}$}
\label{sec:rsp_m}

\begin{figure*}[t]
    \centering
    \includegraphics[width=0.3\textwidth]{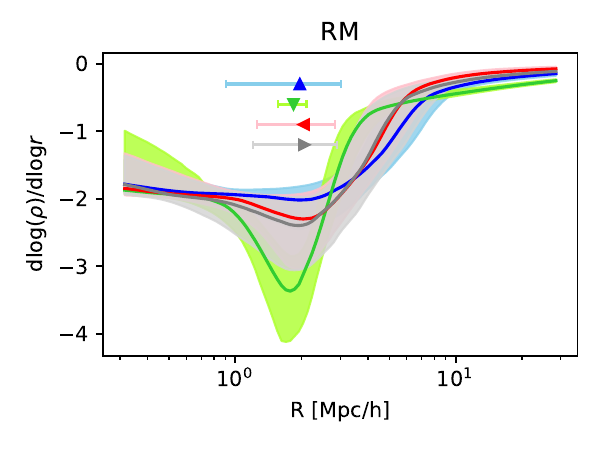}
    \includegraphics[width=0.3\textwidth]{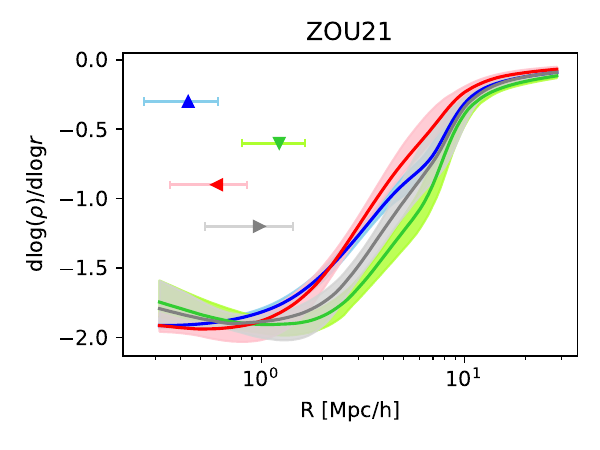}
    \includegraphics[width=0.3\textwidth]{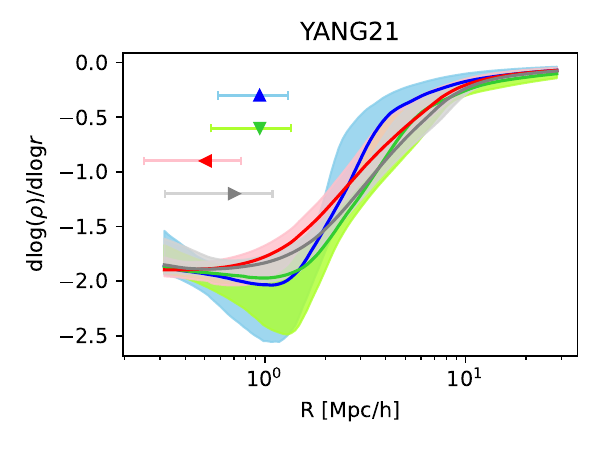}
    \includegraphics[width=0.3\textwidth]{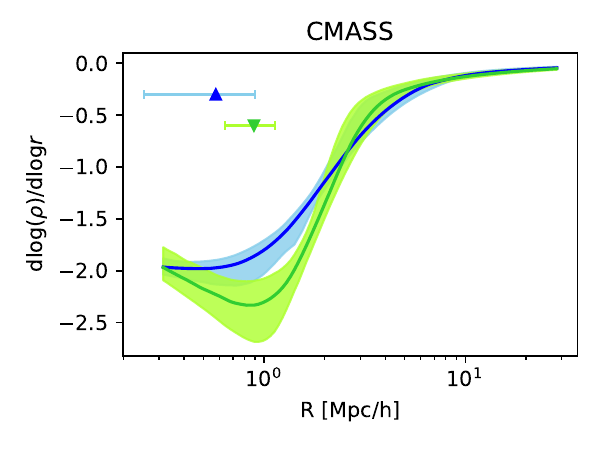}
    \includegraphics[width=0.3\textwidth]{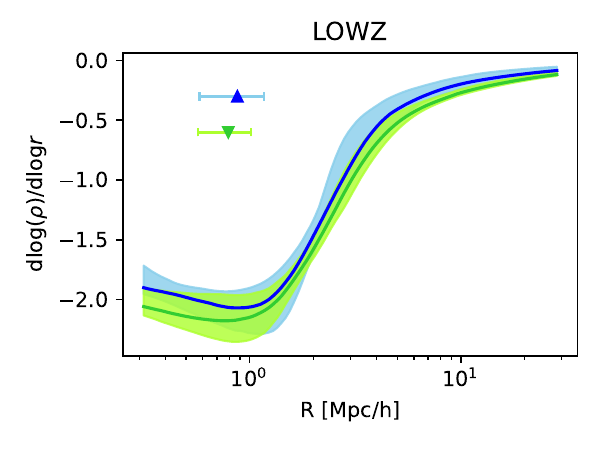}
    \caption{Profile of the density slope and corresponding value of splashback radius. 
    The catalog name is shown at the top of each panel. 
    The colors and symbols are consistent with Fig.~\ref{fig:ggl_signal}. The solid curve and its corresponding light-colored shading represent the best-fit model and its 1$\sigma$ range. The data point and horizontal error bar denote the best-fit splashback radius and its 1$\sigma$ range. We use different $Y$ values of the data points to separate the results of bins.}
    \label{fig:sbr_plot}
\end{figure*}

We investigate the density slope profiles and the associated splashback radii for each catalog bin to gain an insight into the factors that influence the splashback radius (see Fig.~\ref{fig:sbr_plot}). Our findings demonstrate a general positive correlation between splashback radius and the mass of the clusters, while there appears to be no dependence of splashback radius on redshift.

\begin{figure*}
    \centering
    \includegraphics[width=0.98\textwidth]{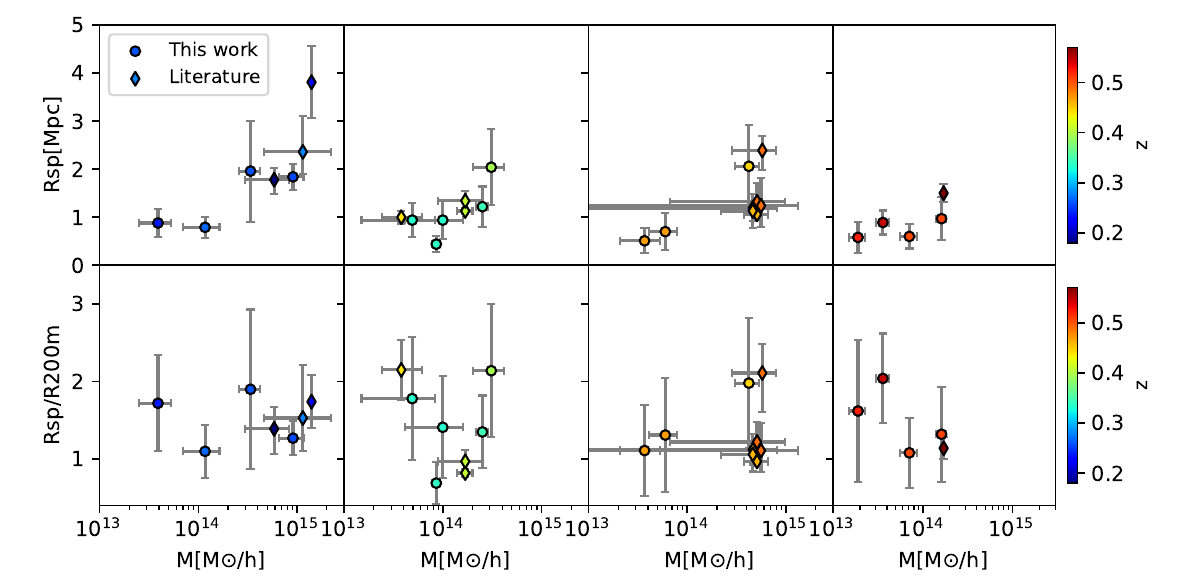}
    \caption{The relation between halo mass and the splashback radius (panels in the first row) or the normalized splashback radius (panels in the second row) in 4 redshift bins. In each row, from left to right, panels show halos from the low redshift to the high redshift. Measurements from this work and literature are shown with dots and diamonds, respectively. The color of the data point represents the redshift.}
    \label{fig:sbr_colorMZ_4zbins}
\end{figure*}

In Fig.~\ref{fig:sbr_colorMZ_4zbins}, the relation between halo mass and the splashback radius or the normalized splashback radius is plotted in 4 redshift bins, with redshift represented by the color of data points. The sample size of each redshift bin is roughly equal.
In each redshift bin, there exists a positive tendency between the splashback radius and halo mass remains, especially in the lowest redshift bin. However, this correlation between them becomes much weaker when the splashback radius is normalized. And the relation shows a U-shape, especially in the two lowest redshift bins. However, the measurement uncertainty is too large for a robust conclusion.

\begin{figure*}
    \centering
    \includegraphics[width=0.98\textwidth]{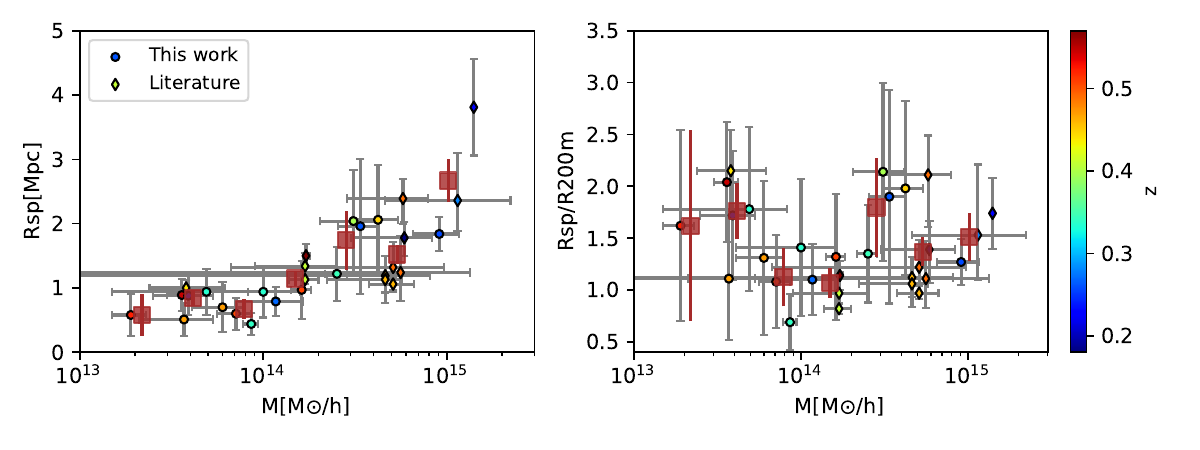}
    \caption{The relation between halo mass and the splashback radius (left panel) or the normalized splashback radius (right panel). Measurements from this work and literature are shown with dots and diamonds, respectively. The color of the data point represents the redshift. The whole sample is rebinned and shown in brown squares with errorbars.}
    \label{fig:sbr_colorMZ}
\end{figure*}

In the left panel of Fig.~\ref{fig:sbr_colorMZ}, we illustrate the correlation between halo mass and the splashback radius for the whole sample across the whole redshift range. The positive correlation between the splashback radius and halo mass remains. Additionally, we depict the relation between the normalized splashback radius and the halo mass in the right panel therein. The U-shaped trend is distinctive, with a pivot point around $10^{14} M_{\odot}$.
However, the upturns in both relations are likely caused by massive halos with low redshifts, because low redshift halos correspond to a lower peak height than high redshift halos with the same mass.
In both plots, there is no clear observable correlation between the splashback radius and redshift at a given halo mass. Notably, our measurements align consistently with literature results within a 1$\sigma$ range.

\subsection{The relationship between the splashback radius and peak height}
\label{sec:rsp_nu}

Theoretically, the peak height of the halo influences the accretion rate and the radial motion of the accreted matter, thus it also impacts the splashback radius. We explore the relation between the peak height and the splashback radius. Following \cite{More2015}, we calculate peak height as 
\begin{equation}
    \nu=\nu_{\rm 200m} \equiv \delta_c / \sigma(M_{\rm 200m})/D(z), 
    \label{eq:nu_z}
\end{equation}
where $\delta_c$ denotes the critical threshold at which the collapse begins, $\sigma^2$ is the fluctuation variance of the initial density map, and $D(z)$ is the growth factor. 

\begin{figure*}
    \centering
   \includegraphics[width=0.95\textwidth]{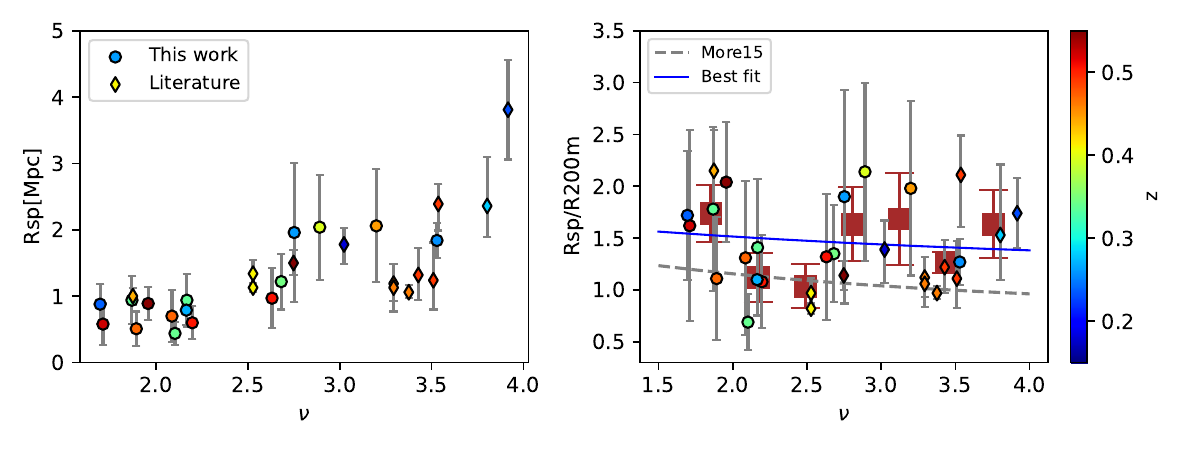}
    \caption{The relation between the peak height and splashback radius (left panel) and the normalized splashback radius (right panel). The symbols are the same with Fig.~\ref{fig:sbr_colorMZ}. The rebinned result of the whole sample is shown with the brown squares with corresponding errorbars. The model from \cite{More2015} is overlaid as the grey dashed curve, while the best-fit model is shown in the blue solid curve.}
    \label{fig:sbr_nu}
\end{figure*}

In Fig.~\ref{fig:sbr_nu}, we depict the relationship between the peak height and the normalized splashback radius.
We find the splashback radius increases with the peak height, although the scatters are large. However, this tendency is also highly affected by massive halos at low redshifts, whose sizes are large in the case of virial radius. 
Thus, the normalized splashback radius is also depicted for its relation with halo mass in this Figure. We find the upturn of splashback radius for massive halos has been greatly reduced after it is normalized. In both panels, our measurements are aligned with literature in $\sim 1\sigma$ range. 
Furthermore, the binned data of the whole sample are compared with the parameterized function from the simulation work of \cite{More2015}, as
\begin{equation}
\begin{split}
    R_{\rm sp}/R_{\rm 200m}=A~(1+B~e^{-\nu/C}).\\
    \end{split}
    \label{eq:rspr200m_nu}
\end{equation}
In the work of \cite{More2015}, the parameters in Eq.~\ref{eq:rspr200m_nu} are estimated as $A=0.81$, $B=0.97$, $C=2.44$. However, we find the measurements deviated from the model using these parameter values. The reduced $\chi^2$ for the binned data of the whole sample is $6.73$. 

We fit the binned data of the whole sample with the Eq.~\ref{eq:rspr200m_nu} and obtain the best-fit parameters as $A=1.20^{+0.22}_{-0.64}$, $B=0.66^{+2.91}_{-1.17}$, $C=3.03^{+5.08}_{-2.74}$.
The best-fitting model is overlaid in Fig.~\ref{fig:sbr_nu}. We find the best-fitting model is consistent with the model of \cite{More2015}, although the normalization has some offset mightly due to systematics, such as different fitting procedures \citep{ONeil2021}.

In \citet{Diemer2014}, the relation between the accretion rate and the normalized splashback radius is
\begin{equation}
    R_{\rm sp}/R_{\rm 200m}=0.54~[1+0.53\Omega_{\rm m}(z)]~(1+1.36~e^{-\Gamma/3.04}),
    \label{eq:gamma}
\end{equation}
where the $\Gamma$ is the accretion rate.
The normalized splashback found in the simulation has a negative correlation with the accretion rate \citep{Diemer2014, Diemer2017}. It is the result of the shrinking apocenter of accreted matter in the rapidly growing potential well.

Specifically, \cite{More2015} constrain the splashback radius from halo merger trees from numerical simulations, and find $R_{\rm sp}$ is $0.8-1R_{200\rm m}$ for halos with rapid accretion and is $\sim 1.5 R_{200\rm m}$ for halos with slow accretion. 
In the right panel of Fig.~\ref{fig:sbr_nu}, we find the splashback radius is mostly $\gtrsim$ $R_{\rm 200m}$ in the whole sample, meaning these halos tend to have a low accretion rate.

Furthermore, several other theoretical studies are exploring the dependence of the splashback radius on parameters such as mass accretion rate ($\Gamma$), height peak ($\nu$), and redshift ($z$). Using a simple spherical collapse model of cosmological N-body halos, \cite{Adhikari2014} concluded that the normalized splashback radius ($R_{\rm sp}$/$R_{\rm 200m}$) is smaller for halos with higher accretion rates, where particle orbits contract within deep potential wells, particularly for halos located in low-redshift regions with low background density. This finding is supported by \cite{Shi2016}, who suggested that the correlation with redshift originates from a high $\Omega_{\rm m}$ and the accretion rate at high redshift.
Although \cite{Diemer2014} derived the normalized splashback radius from the density profiles of $\Lambda$CDM halos and found it to be small for halos with high accretion rates but with no redshift evolution, the recent works in \cite{Diemer2017,Diemer2020} find some evidence of the redshift evolution of $R_{\rm sp}$.

\section{Conclusion}
\label{sec:conclusion}

In this study, we measure and model the combined weak lensing signal for the redMaPPer, ZOU21, YANG21, CMASS, and LOWZ catalogs, which are divided into different redshift and mass.
We employ a three-component model to fit $\Delta\Sigma$(R), including contributions from the central halo, miscentering effect, and the external halo. The splashback radius for each bin is derived from the best-fit $\Delta\Sigma$(R). We highlight the broad mass and redshift ranges over which we measure the splashback radius.

Our findings indicate a consistent increase in the splashback radius with halo mass across the entire mass range. However, the normalized splashback radius $R_{\rm sp}/R_{\rm 200}$ does not exhibit a monotonic growth with halo mass; instead, it displays a U-shaped trend, which might result from the redshift evolution.
In addition, we find a positive relation of the splashback radius and a negative relation of the normalized splashback radius with the peak height, and the latter one is consistent with the prediction from \cite{More2015}.
Furthermore, in the whole sample, combining literature and our measurements, the splashback radius is mostly $\gtrsim R_{\rm 200m}$, meaning these halos tend to have a low accretion rate.
However, the uncertainty in the measurements is very large.
Future, more in-depth observations, and simulations are crucial for resolving this issue.

\software{Cluster~toolkit \citep{Smith2003,Eisenstein1998,Takahashi2012}, 
Colossus software \citep{Diemer2018}, 
CAMB \citep{Challinor2011,Lewis1999}, 
SWOT \citep{Coupon2012}, 
EMCEE \citep{emcee}, 
WebPlotDigitizer \citep{Rohatgi2020}.}

\begin{acknowledgements}
We would like to thank the anonymous referee for detailed comments and suggestions, resulting in an improved version of the manuscript.
This work is supported by National Key R$\&$D Program of China No. 2022YFF0503403 and the Ministry of Science and Technology of China (grant Nos. 2020SKA0110100). We acknowledge the support of National Nature Science Foundation of China (Nos 11988101,11773032,12022306,12203063), the support from the Ministry of Science and Technology of China (grant Nos. 2020SKA0110100), the science research grants from the China Manned Space Project (Nos CMS-CSST-2021-B01,CMS-CSST-2021-A01), CAS Project for Young Scientists in Basic Research (No. YSBR-062), and the support from K.C.Wong Education Foundation. WX thanks Rui Wang for useful statistical discussion during the work. HYS acknowledges the support from NSFC of China under grant 11973070, Key Research Program of Frontier Sciences, CAS, Grant No. ZDBS-LY-7013 and Program of Shanghai Academic/Technology Research Leader. JY acknowledges the support of the National Science Foundation of China (12203084), the China Postdoctoral Science Foundation (2021T140451), and the Shanghai Post-doctoral Excellence Program (2021419). CZ acknowledges support from Zhejiang Provincial Natural Science Foundation of China under grant No. LQ24A030001.
\end{acknowledgements}

\bibliography{main} 
\bibliographystyle{aasjournal}

\end{document}